\documentclass{ws-ijmpcs}
\newcommand{\be}{\begin{equation}}
\newcommand{\ee}{\end{equation}}
\newcommand{\ba}{\begin{eqnarray}}
\newcommand{\ea}{\end{eqnarray}}
\newcommand{\non}{\nonumber}
\def\cF{{\cal F}}
\def\cG{{\cal G}}
\def\dce{\varepsilon_{\cal G}'}

\def\d{\partial}
\def\e{{\rm e}}
\def\cE{{\cal E}}
\def\cH{{\cal H}}
\newcommand{\ep}{\epsilon}
\newcommand{\Vi}{V_{\rm int}} 
\newcommand{\lb}{\label}
\newcommand{\p}{\partial}

\newcommand{\ve}{\varepsilon}
\newcommand{\eps}{\epsilon }
\newcommand{\al}{\alpha}

\newcommand{\la}{\lambda}
\newcommand{\Bk}{\ensuremath{\Sigma}}
 
\begin{document}

\markboth{D.V. Gal'tsov and E.A. Davydov} {Yang-Mills condensates in
cosmology}

%
\catchline{}{}{}{}{}
%

\title{YANG-MILLS CONDENSATES IN COSMOLOGY }

\author{D.V. GAL'TSOV}

\address{Department of Theoretical Physics, Moscow State
University,\\ 119899, Moscow, Russia\\
galtsov@physics.msu.ru}
\author{E.A. DAVYDOV}

\address{Bogoliubov
Laboratory of Theoretical Physics, JINR, \\141980, Dubna, Moscow
region, Russia\\
davydov@theor.jinr.ru}

\maketitle


\begin{abstract}
We discuss homogeneous and isotropic cosmological models driven by
$SU(2)$ gauge fields in the framework of Einstein gravity. There
exists a  Yang-Mills field configuration, parametrized by a single
scalar function,  which consists of parallel electric and magnetic
fields and has the stress tensor mimicking  an homogeneous and
isotropic fluid. The unique $SU(2)$ gauge theory with spontaneous
symmetry breaking sharing the same property is the Yang-Mills
coupled to the complex doublet Higgs, this exists only in the case
of the closed universe. This model contains an intrinsic mechanism
for inflation due to the Higgs potential. Our second goal is to show
that a successful inflation can be achieved also within the pure
Yang-Mills theory adding an appropriate theta-term. 
\end{abstract}

\ccode{PACS numbers: 11.25.Hf, 123.1K}

\section{Introduction}
A remarkable fact is that the $SU(2)$ Yang-Mills (YM) field admits
isotropic and homogeneous configurations parameterized by a single
scalar function. This is somewhat surprising since the
energy-momentum tensor of a single  vector field in flat spacetime
always contains anisotropic stresses $E_iE_j$, $B_iB_j$ so this
feature is clearly impossible in the $U(1)$ case without spatial
averaging. In the $SU(2)$ case one has three vector potentials
$A_\mu^a$ producing electric and magnetic component fields
$E_i^a,\;B_i^a$ which make the stress-tensor diagonal and isotropic
if $E_i^a=E\delta_i^a$ and $B_i^a=B\delta_i^a$, indeed
$E_iE_j=E^2\delta_i^a \delta_j^a=E^2\delta_{ij}$. The electric and
magnetic vectors are parallel, so both field invariants, the scalar
$\cF=F^a_{\mu\nu}F^{a\mu\nu}/2=3(B^2-E^2)$ and the pseudoscalar
$\cG=\eps^{\mu\nu\la\tau}F^a_{\mu\nu}F^a_{ \la\tau}/8=3EB$ are
non-zero.

Such an ansatz can be  generalized to any of the FRW metrics \be
ds^2= dt^2-a(t)^2 [d\chi^2+\Bk_k(d\theta^2+\sin^2\theta d\phi^2)],
\ee closed ($k=1,\,\Bk_1=\sin  \chi$), open ($k=-1,\, \Bk_{-1}=\sinh
\chi$) or spatially flat ($k=0 ,\, \Bk_{0}=  r$) and in the matrix
form reads~\cite{Galtsov:1991un}
\begin{eqnarray}\label{Fform}
F&=&F^aT^a=\dot{f}\left(T_n\,dt\wedge d\chi
     + T_\theta \Bk_k\,dt \wedge d\theta
     + T_\phi \Bk_k \sin \theta \,dt \wedge d\phi \right) \nonumber \\
     &&+ \Bk_k(f^2-k)\left(T_\phi\,d\chi \wedge d\theta
     - T_\theta \sin \theta\,d\chi \wedge d\phi
     + T_n \Bk_k  \sin \theta\,d\theta \wedge d\phi \right).
\end{eqnarray}
Here the rotating $SU(2)$ generators are used: \be T_n=\tau^a
n^a/2i,\;T_\theta=\tau^a e_\theta^a/2i,\;T_\phi=\tau^a
e_\phi^a/2i,\ee where $n^a,\,e_\theta^a,\,e_\phi^a $ are the
spherical unit vectors, and $\tau^a$ are Pauli matrices. This
property remains valid also for larger gauge groups containing an
embedded SU(2)~\cite{Moniz:1990hf,Darian:1996mb}. Remarkably, the
EYM cosmology is also solvable at the quantum level,  the quantum
FRW cosmology having been discussed in a number of papers in the
90-ies~\cite{Cavaglia:1993en}. An interesting feature of the EYM
quantum cosmology is a possibility of tunneling transitions between
de Sitter and hot FRW cosmologies \cite{worm}. Similar models were
formulated in the context of supergravities \cite{Moniz:1996ja}. In
the spirit of the string theory the Born-Infeld generalizations were
also considered \cite{BI}. More recently, the YM cosmological models
with modified lagrangians were suggested in the context of the dark
energy models~\cite{YMDE,1,Maleknejad:2011sq}.

Now consider the gauge theory with spontaneous symmetry breaking.
For the $SU(2)$ group one has two basic possibilities for the
coupled Yang-Mills-Higgs (YMH) system: the real triplet Higgs  and
the complex doublet Higgs. The non-interacting Higgs belonging to
any representation of the gauge group can form isotropic and
homogeneous configuration. But for an interacting YM-Higgs system it
is not so. It turns out that no homogeneous and isotropic
configuration of interacting YM-triple Higgs  exist which would be
compatible with the Friedman-Robertson-Walker metric for any value
of $k$. In the complex doublet case such a configuration exists in
the closed case $k=1$ \cite{Gal'tsov:2010dd}. This has a simple
explanation: The doublet Higgs has $SO(4)$ symmetry which is the
isometry group of the three-sphere. This is another interesting
coincidence, which selects the YM-doublet Higgs as natural
``realistic'' field-theoretical model for cosmology. This model can
accommodate for inflation, and at the same time naturally introduces
a vector field which could play a role of curvaton
\cite{Dimopoulos:2009am}.

\section{Yang-Mills-Higgs cosmology}
Consider  the EYMH action with complex doublet Higgs:
 \be
    S=
     \int \left\{-\frac{1}{16\pi G}R
      -\frac14  F^a_{\mu\nu}F^{a\mu\nu} +
    \frac{1}{2}(D_\mu \Phi)^\dag D_\mu \Phi-\frac{\lambda}{4}
    \left(\Phi^\dag\Phi-v^2\right)^2\right\}\sqrt{-g}d^4x,
 \ee
where $\Phi$ is the complex doublet Higgs, and $
D_\mu\Phi=\partial_\mu\Phi+gA^a_\mu T_a\Phi. $ The YM  matrix
potential $A_\mu=A_\mu^a T_a$ generating the field strength
(\ref{Fform}) for $k=1$ can be chosen as
 \be \label{amu}
       A_\mu^a T_a=\frac{1-f(t)}{2g}\,U\,\partial_\mu U^{-1},
       \quad U=\e^{2\chi T_n },
 \ee
 while an
ansatz for Higgs is
 \be \label{hig}
      \Phi=h(t) U\Phi_0,
      \quad \Phi_0^\dag\Phi_0=1,
  \ee
  where $\Phi_0$ is an arbitrary normalized constant spinor.
The compatibility is this configuration with the required spatial
symmetries follows from the identity
 \be
    D_\mu\Phi=\frac{1+f(t)}2\partial_\mu\Phi.
\ee
  The EYMH action contains three different mass parameters: the
Planck mass, the mass of the W-boson, and the Higgs mass
 \be
 M_{\rm Pl}=\frac1{\sqrt{G}},\qquad M_W=gv,\qquad M_H=\sqrt{\la} v.
 \ee
We rescale the Higgs scalar as $h\to h M_{\rm Pl}$ and introduce
dimensionless parameters
 \be
\alpha=\frac{M_W}{gM_{\rm Pl}}=\frac{v}{M_{\rm Pl}},\qquad
\beta=\frac{M_H}{M_W}=\frac{\sqrt{\la}}{g}.
 \ee
Finally, introducing the reduced Planck length  $l=1/(g M_{\rm
Pl})$, we present the metric in dimensionless terms:
 \be
 ds^2=l^2\left\{-N^2 dt^2+a^2\left[d\chi^2+\sin\chi^2
 (d\theta^2+\sin^2\theta d\varphi^2)\right]\right\}.
  \ee
Substituting the ansatze (\ref{hig},\ref{amu}), integrating over
angles and dividing by the volume of the unit  the three-sphere,  we
obtain the one-dimensional reduced action $S = S_{\rm g}+S_{\rm
YM}+S_{\rm H}+S_{\rm int},$ where
 \ba
S_{\rm g} &=&\int
\frac3{8\pi}\left(aN-\frac{\dot{a}^2a}{N}\right)dt,\quad
S_{\rm YM}=\frac32\int\left(\frac{ \dot{f}^2 a}{ N}-\frac{ N(f^2-1)^2}{ a}\right)dt,\\
S_{\rm H}&=&\int \left(\frac{\dot{h}^2 a^3}{2N}
-\frac{\beta^2}4(h^2-\alpha^2)^2Na^3\right)dt,\;\;
 S_{\rm int}=-\frac34\int
h^2(f+1)^2 Na dt,
 \ea
and the total derivative in the gravitational term is omitted.

 Variation with respect to $N$  leads to the constraint
 equation
\begin{equation}\label{eq:N}
     \left(H^2+\frac1{a^2}\right)=\frac{8\pi }{3}\ve,\quad
     H=\frac{\dot{a}}{a},
\end{equation}
and we fix the gauge $N=1$ afterwards. For the energy density we
 obtain:
\begin{equation}\label{eq:rho}
    \ve=\frac12\dot{h}^2+\frac{3 \dot{f}^2}{2 a^2}+V_f+V_h+\Vi,
\end{equation}
where the  potentials read:
 \be
 V_f=\frac{3(f^2-1)^2}{2a^4},\quad
V_h=\frac{\beta^2}4(h^2-\alpha^2)^2 ,\quad V_{\rm int}=\frac{3
h^2(f+1)^2 }{4a^2}.
 \ee
The acceleration equation is obtained by variation of the action
over $a$ with account for the constraint:
\begin{equation}\label{eq:a2}
     \frac{\ddot{a}}{a}=-\frac{4\pi }{3}(\ve+3p),
\end{equation}
leading to the following expression for pressure:
\begin{equation}\label{eq:p}
    p=\frac12\dot{h}^2+\frac{ \dot{f}^2}{2 a^2}+\frac13 V_f-V_h   ,
\end{equation}
and therefore,
\begin{equation}\label{eq:a3}
    \frac{\ddot{a}}{a}=-\frac{8\pi }{3}\left(\dot{h}^2 +\frac{3 \dot{f}^2}{2 a^2}
    -V_h+V_f\right).
 \end{equation}
Note that $V_h$ and $V_f$ enter with different signs, while the
interaction term $V_{\rm int}$ does not enter the acceleration
equation at all. The field equations for the YM and Higgs scalar
functions read
\begin{eqnarray}
  \ddot{h} + 3H\dot{h} &=&
   -\frac{3}{2a^2}h(f+1)^2-
  \beta^2(h^2-\al^2)h , \non\\
  \ddot{f}+H\dot{f} &=&-\frac{3}{2}h^2(f+1)-\frac{6}{ a^2}(f^2-1)f.\label{eq:w}
\end{eqnarray}

It is worth noting, that two roots $f=\pm 1$ of the YM potential
$V_f$ correspond to two neighboring topological vacua. To show this,
consider the Chern-Simons 3-form
 \be
 \omega_3=\frac{g^2}{8\pi^2} {\rm Tr} \left(
A\wedge dA -\frac{2ie}3 A\wedge A\wedge A\right),
 \ee satisfying the equation
  \be
d\omega_3= \frac{e^2}{8\pi^2} {\rm Tr} F\wedge F.
 \ee
Substituting here the  potential (\ref{amu}) one finds that it is
non-trivial in the closed universe $k=1$, giving the winding number
of the map $SU(2)\to S^3$:
 \be
  N_{CS}=\int_{S^3} \omega_3=\frac14
(f+1)^2(2-f).
 \ee
 The vacuum $f=-1$ is therefore topologically trivial: $N_{CS}=0$,
while the vacuum $f=1$ is the next non-trivial one with $N_{CS}=1$.
This does not mean, however, that both will be classical solutions
to our system.

One can see that the above equations have the following particular
solutions:
\begin{itemize}
\item If Higgs is in the false vacuum  $h=0$, we have the EYM
system with the cosmological constant $\Lambda=\alpha^4\beta^2/4$;
the corresponding scale factor can be found in a closed form
\cite{Gal'tsov:2010dd}. The Euclidean sector of the same theory
contains instanton and wormhole solutions  interpolating between the
de Sitter and the hot universes \cite{Donets:1992ck}.
\item A particular (unstable) analytic solution is cosmological
sphaleron in the false Higgs vacuum: \vspace{-.2cm}\be\lb{fp1}
h=0,\quad f=0,\quad a^2=\frac{\sqrt{6}}{\beta\alpha^2} .\ee This
solution describes the YM field sitting at the top of the potential
barrier separating two topologically neighboring YM vacua (note that
$f=0$ is not vacuum for YM ) and generalizes the pure EYM sphaleron
of Ref.~\refcite{Gist}. In this case $H=0$ and the universe is in
the steady state.
\item There is another non-trivial steady state with non-zero $f,\;h$:
\be\lb{fp2} f=\frac{\sqrt{8}\beta -\sqrt{3}}{\sqrt{8}\beta
+\sqrt{3}},\quad h=\sqrt{f}\alpha,\quad
a^2=\frac{4(1-f)}{\alpha^2}.\ee
\item If YM  field is in the trivial vacuum $f=-1$,
we have a pure  Higgs $k=1$ cosmology, which is the basis of the
standard inflationary models.
\end{itemize}
At the same time, there are several regimes which could be expected,
but actually are not realized as classical solutions of the EYMH
system:
\begin{itemize}
\item The system
never reduces to the pure EYM cosmology with the conformal equation
of state $p=\frac13 \ve$ and the hot universe metric as described in
Ref.~\refcite{Galtsov:1991un}. \item The non-trivial vacuum $f=1$ is
not a solution to our system: the interaction term does not vanish
in this case, so the YM equation will not be satisfied. Thus, the YM
field dynamically coupled to Higgs will have the unique vacuum state
$f=-1$.\item The Higgs vacuum $h=\alpha$ is not a solution unless
the YM field is vacuum $f=-1$.
\end{itemize}
The system possesses the slow-roll regime when kinetic terms
$\dot{f},\; \dot{h}$ are small relative  to potential terms. There
are three different slow-roll regimes:
\begin{itemize}
\item  inflation, with an approximate equation of state  $p=-\ve$
in the case of the dominant Higgs potential $V_h$; \item hot
universe with $p= \ve/3$ in the case of dominant YM ``potential''
$V_f$; \item   zero-acceleration regime with the equation of state
$p= -\ve/3$  in the case of dominant interaction term $V_{\rm int}$.
The emergence of such an equation of state (usually referred as the
string gas equation) regime  is somewhat unexpected, it also occurs
in the high energy limit of the non-Abelian Born-Infeld cosmology
\cite{BI}.
\end{itemize}
Note, that $V_f$ and $V_{\rm int}$ are not really the potentials:
they depend on the cosmological scale factor.

The solutions (\ref{fp1}) and (\ref{fp2}) are fixed points of the
six-dimensional dynamical system associated with the equations of
motion. The first one corresponds to the local maximum of the Higgs
potential, in which case interaction between the Higgs and the YM
fields is switched off. The second point corresponds to the minimum
of the potential $h_{\rm min}<\alpha$, shifted by interaction with
YM. Correspondingly, the position of the YM function $f$ is also
shifted from the extrema of $V_f$. In this case the interaction
plays crucial role. The analysis of perturbations around the fixed
points shows that the first one is an unstable node. For the second
fixed point one obtains stable and unstable directions in the phase
space depending on the value of the parameter $\beta$, see
Ref.~\refcite{Gal'tsov:2010dd} for details. The most interesting
case corresponds to stability of this point against small
perturbation, which translates in the existence of long quasistable
states which terminates by an inflationary period. The number of
$e$-folds during inflation depends on the initial value of the Higgs
field in the quasistationary phase.

Qualitatively, behavior of the integral curves obtained in numerical
experiments \cite{Gal'tsov:2010dd} is somewhat reminiscent of the
hybrid inflation with two truly scalar fields. The interaction
potential $V_{\rm int}$ plays the role of trapping potential which
drives the inflaton to climb the Higgs potential. As a result, the
YM field can substantially enhance the scalar inflation.

\section{Theta-inflation} Consider now the pure YM theory described
by the action $S=\int L \sqrt{-g} d^4x$, where the lagrangian
$L(\cF,\cG)$ depends  arbitrarily  on  two YM invariants
 \be \cF=-F^a_{\mu\nu}F^{a\mu\nu}/2,\quad
\cG=-\tilde F^{a\mu\nu}F^a_{\mu\nu}/4,\quad \tilde F^{a\mu\nu}=
\frac{\eps^{\mu\nu\lambda\tau}F^a_{\lambda\tau}}{2\sqrt{-g}}.\ee We
will focus here on the  dependence of the lagrangian on the
pseudoscalar $\cG$. In gauge theories the linear in $\cG$ term is
induced by instantons and is called theta-term. We will assume,
however, more general dependence of $L$ on $\cG$, motivated, e.g. by
vacuum polarization.

In the Refs.~\refcite{1}, \refcite{Gal'tsov:2010dd} we derived the
equation of state for FRW cosmologies driven by the YM field
configuration (\ref{Fform}) for a general lagrangian $L(\cF,\cG)$.
The contribution coming from $\cG$ was found to produce the energy
density and pressure with the ratio $w=\ve/p=-1$. It is not
difficult to show that this does not depend on the particular
configuration of the YM field. The linear functional $S_{\cG}=\int
\cG \sqrt{-g} d^4x$ does not depend on the metric:
 \be
S_{\cG}=-\frac12\int\frac{\eps^{\mu\nu\lambda\tau}}{\sqrt{-g}}
F_{\mu\nu}F_{\la\tau}\sqrt{-g} d^4x=-\frac12\int
\eps^{\mu\nu\lambda\tau}  F_{\mu\nu}F_{\la\tau} d^4x
 \ee
and hence does not contribute to the energy-momentum tensor.
However, for more general dependence of the lagrangian on $\cG$, it
is not so, since \vspace{-.1cm}
 \be
\frac{\d \cG}{\d g^{\mu\nu}}=\frac12\cG g_{\mu\nu}.
 \ee
Using this, we obtain  the   YM energy-momentum tensor in the form
 \be
T_{\mu\nu}= \frac{2}{\sqrt{-g}}\frac{\delta S}{\delta
g^{\mu\nu}}=2\frac{\d L}{\d \cF}\frac{\d \cF}{\d
g^{\mu\nu}}+\left(\frac{\d L}{\d \cG}\cG-L\right)g_{\mu\nu},
 \ee
where the second term looks like the variable cosmological constant.
The field invariants for the YM configuration (\ref{Fform}) read \be
\cF=\frac{3\dot{f}^2}{a^2N^2}-\frac{3}{a^4}(k-f^2)^2,\quad
\cG=\frac{3\dot{f}(k-f^2)}{Na^3}. \ee We also introduce  the
effective electric and magnetic fields  \be \cE=\frac{\dot{f}}{a
N},\quad \cH=\frac{k-f^2}{a^2} \quad\mbox{so that}\quad
\cF=3(\cE^2-\cH^2),\quad \cG=3\cE\cH. \ee Then in the context of the
FRW models we find for the energy density and pressure:
  \ba
\ve&=&-L +3\cE\left(2\cE\frac{\partial L}{\partial \cF}
+\cH\frac{\partial L}{\partial \cG}\right),\non\\
  p&=&L +2\left(2\cH^2-\cE^2\right)\frac{\partial
L}{\partial \cF} - 3\cE\cH\frac{\partial L}{\partial \cG}. \ea

Consider now the lagrangian with a ``potential'' depending on $\cG$:
 \be
L(\cF,\cG)=\frac{\cF}{2}-V(\cG).\ee  Then \vspace{-.5cm}
\ba \ve_{\cF} &=&\frac{3}{2}(\cE^2+\cH^2),\quad \ve_{\cG}=V-3\cE\cH V',\non\\
\ve &=&\left(\ve_{\cF}+\ve_{\cG}\right),\quad p
=\left(\frac{\ve_{\cF}}{3}-\ve_{\cG}\right), \ea and the
Friedman equations  read: \ba H^2+\frac{k}{a^2}&=&\frac{8\pi}{3} \left(\ve_{\cF}+\ve_{\cG}\right)\non\\
\frac{\ddot a}{a} =\dot
H+H^2&=&\frac{8\pi}{3}\left(\ve_{\cG}-\ve_{\cF}\right).\ea  The
dynamics of the YM field in this model is governed by the equation:
\be \frac{d}{dt}\left(a^3\frac{\p\cF}{\p\dot
f}-a^3V'\frac{\p\cG}{\p\dot f}\right)=a^3\left(\frac{\p\cF}{\p
f}-V'\frac{\p\cG}{\p\dot f}\right).\ee Using the relation $
V''=-\frac{\dce}{3\cE\cH} $, where primes denote derivatives with
respect to $\cG$, after some rearrangements we obtain:  \be
\left[1+\dce \frac{\cH}{\cE}\right]\ddot f +\left[1-3\dce
\frac{\cH}{\cE}\right]H\dot f=2f\cH\left[1+\dce
\frac{\cE}{\cH}\right].\label{eq:YMmot2}\ee We will investigate the
regime when $\dce$ terms dominates in the squared brackets, then the
Eq.~(\ref{eq:YMmot2}) is simplified to \be \ddot f-3 H\dot f
=2f\frac{\cE^2}{\cH}.\ee

Contrary to the scalar inflation generated by the potential term,
here the desired regime rather corresponds to  the kinetic
(electric) term being of the same order as the potential term
(magnetic). Indeed, the theta-term, whose dominance produces
inflation,  is the product of the electric and magnetic components.
The dynamics will depend  on the initial state of the system,
characterized by the ratio $\cH/\cE$, which should not be too small
or too large. Let us investigate the slow-roll regime, which is
expected if
 \begin{itemize}
    \item   Hubble friction  term in the YM equation is dominant;
    \item  the  accelerating term  in the Friedman
    equations is dominant: $\ve_{\cF}\ll \ve_{\cG}$;
    \item  the Hubble parameter is varying slowly,
    $\ep\equiv-\dot{H}/H^2\ll 1,~\eta\equiv -\ddot H/(2H\dot H)\ll 1$;
    \item the curvature term is negligible $k/a^2\ll H^2$.
\end{itemize}
In what follows we set $k=0$, just noting that the case $k=1$ is
rather different and very interesting, it will be considered
elsewhere. For the slow-roll approximation it may be not correct to
use just $\ddot f\ll 3H\dot f$.  It will be better to work with the
ratio $\psi=f/a$, which is connected with the magnetic component via
$\cH=-\psi^2$ and whose derivative is related to  the electric one:
$\cE=\dot\psi+H\psi$. Thus we arrive at the system \ba
&\ddot\psi+2H\dot\psi=-\psi\left(\dot H+
2\frac{\dot\psi^2}{\psi^2}\right),&\non\\
&H^2  = \frac{8\pi}{3} \left(\ve_{\cF}+\ve_{\cG}\right),\quad \dot
H+H^2=\frac{8\pi}{3}\left(\ve_{\cG}-\ve_{\cF}\right).&
\label{eq:sysslow}\ea The dominance of the theta-terms
$\ve_\cG\gg\ve_\cF$ implies smallness of the first slow-roll
parameter: \vspace{-.2cm} \be
\ep=\frac{2\ve_\cF}{\ve_\cF+\ve_\cG}\ll 1. \ee

To simplify further calculations we  proceed with the potential \be
\label{eq:Vln} V(\cG)=-\theta\,\cG\ln\cG,\ee in which case the
theta-energy density  $ \ve_\cG=\theta\cG $ is linear in $\cG$, the
same as the standard $\cF$ term. Then the condition
$\ve_\cF\ll\ve_\cG$ of the accelerated expansion can be  satisfied
in a wide range of  initial conditions and will persist long enough
if  $\theta\gg 1$. In this case the condition $\dce=\theta\gg 1$ is
satisfied and we  arrive at the system (\ref{eq:sysslow}) admitting
an accelerated solution if the electric and magnetic components are
of the same order.

To ensure slow variation of  the Hubble parameter $H$ and the
effective inflaton  $\psi$ during the inflation one needs slowness
of the parameter $\delta=-\frac{\dot\psi}{H\psi}$, in terms of which
the electric and the magnetic components read \be
\cE=(1-\delta)H\psi,\quad
-\frac{\dot\cE}{H\cE}=\frac{\dot\delta}{H(1-\delta)}+\delta+\eps,\quad
-\frac{\dot\cH}{H\cH}=2\delta.\ee By virtue of  the Friedman
equations, $\dot H\sim \ve_\cF$, so we can rewrite the system
(\ref{eq:sysslow}) as \begin{align} &\frac{\dot\delta}{H} =
-2\delta-\eps+\delta\eps+3\delta^2,\quad
\eps  = 2\left[1-\frac{\theta\cE\cH}{\cE^2+\cH^2}\right]^{-1},\non\\
&\eta=\frac{1}{\cE^2+\cH^2}\left[\cE^2
\left(\frac{\dot\delta}{H(1-\delta)}+\delta+\eps\right)+
2\cH^2\delta\right]. \label{eq:slowroll}\end{align} To ensure slow
variation of $\delta$, $\dot\delta/(H\delta)\ll 1$, we set
$\delta=-\eps/2+O(\eps^2)$. Then the last equation in the leading
order in $\eps$ reduces to \vspace{-.1cm}\be
\eta=\frac{\eps}{2}\frac{\cE^2-2\cH^2}{\cE^2+\cH^2}, \ee therefore
$\eta\sim\eps,~\dot\eps\sim H\eps^2$. Finally,  smallness of $\eps$
implies the constraint on the initial state and the parameter
$\theta$: \vspace{-.2cm}\be \frac{\cE^2+\cH^2}{\theta\cE\cH}\ll
1,\label{eq:eps}\ee so $\theta$ should be large enough, with both
electric and magnetic fileds  non-vanishing.

The number of $e$-folds is given by \be \label{eq:Ne1}
N_e=\int_{t_i}^{t_e}Hdt=-\int_{H_i}^{H_e}\frac{d H}{\eps H}.\ee
Using the relation between variations of  $\psi$ and $H$ due to
 $\delta=-\eps/2$, we obtain \be
\frac{\psi}{\psi_i}=\sqrt{\frac{H_i}{H}},\label{eq:psiHubble}\ee
where the initial values must satisfy the constraint \be
H_{i}^2\simeq \frac{8\pi}{3}\ve_\cG\simeq 8\pi\theta
H_i\psi_{i}^3,\quad\Rightarrow\quad
H=\frac{8\pi\theta\psi_{i}^5}{\psi^2}.\ee The slow roll ends when
the condition (\ref{eq:eps}) fails. During the inflation the Hubble
parameter decreases since $\dot H=-16\pi\ve_\cF/3$, therefore the
inflaton $\psi$ goes up. The electric component decreases as
${\cE\simeq H\psi\sim 1/\psi}$, while the magnetic increases as
$|\cH|\sim \psi^2$. The slow roll regime stops when the ratio
$\cH/\cE$ approaches the value $\theta$, i.e. the imbalance between
the electric and the magnetic components becomes comparable with the
imbalance of the standard and $\theta$-term in the lagrangian. This
happens when  \be \frac{\cH(\psi_e)}{\cE(\psi_e)}\sim
-\theta\quad\Rightarrow\quad\psi_{e}^3\sim
8\pi\theta^2\psi_{i}^5.\label{eq:psi_e}\ee Then the dominance of the
$\dce  {\cE}/{\cH}$ term at the right hand side of the
Eq.~(\ref{eq:YMmot2}) ends, and the system dynamics changes.

To estimate the number of $e$-folds, we substitute
$dH/H=-2d\psi/\psi$ into (\ref{eq:Ne1}) using the
Eq.~(\ref{eq:psi_e}) to relate   $\phi_e$ and $\phi_i$. For  $\eps$
use the system (\ref{eq:slowroll})  expressing it in terms of $\psi$
alone via (\ref{eq:psiHubble}). During the main part of the
inflation stage one has $\cH>\cE$, so that $1/\eps\simeq
-\theta\cE/(2\cH)=\theta H_i\psi_{i}^2/(2\psi^3)$. Finally we
obtain: \be N_e\approx \theta
H_i\psi_{i}^2\int_{\psi_i}^{\psi_e}\frac{d\psi}{\psi^4}= \frac{
8\pi\theta^2\psi_{i}^5}{3}\left(\frac{1}{\psi_{i}^3}-\frac{1}{\psi_{e}^3}\right)=\frac{
8\pi\theta^2}{3}\psi_{i}^2-\frac{1}{3}.\ee  Unlike the scalar case
when the scalar field is rolling \textit{down}, the YM inflaton
$\psi$ is rolling \textit{up} during the inflation. To avoid going
beyond the Planck mass we should  properly choose the initial
conditions. Taking as the value of $\psi_e$ the rescaled Planck mass
$g M_{\rm Pl}$, we obtain from (\ref{eq:psi_e}) for the initial
value in the same units: $ \psi_i=(8\pi\theta^2)^{-1/5}\ll
\phi_e\sim 1. $ Hence the number of $e$-folds will be \be
N_e\approx\frac13(8\pi\theta^2)^{3/5}\approx 2.3 \,\theta^{6/5}.\ee
Therefore we need the parameter $\theta$ to be just the order of ten
to obtain the slow roll regime with sufficient number of  $e$-folds.
The strongest inflation takes place when electric and magnetic YM
components initially are of the same order.

\section{Outlook}
Incorporating gauge field into cosmological setting seems natural
and promising. We have shown that homogeneity and isotropy selects
the YM-doublet Higgs as the unique gauge theory with spontaneous
symmetry breaking compatible with the closed FRW cosmology which
admits both the inflationary and the hot universe regimes. Our
second proposal consists in using a suitable theta-term to achieve
cosmic acceleration with the pure YM lagrangian.

\section*{Acknowledgments}
D.G. thanks the organizers of QFEXT-11 for invitation and support,
and A.~Starobinsky, V.~Mostepanenko, C.~Schubert and S.~Gavrilov for
useful discussions. This work was supported by RFBR grants
11-02-01371-a and 11-02-01335-a.


\end{document}